\documentstyle[12pt,epsf]{article}
\begin{document}

\par\vspace{4mm}\par
\begin{center}
 
{\Large \bf Magnon scattering processes \\
and low  temperature resistivity \\
 in CMR manganites}

\par\vspace{4mm}\par
 Nobuo Furukawa$^a$, Yoshihiro Shimomura$^a$, T. Akimoto$^b$ and
 Y. Moritomo$^c$

\sl
\par\vspace{4mm}\par
 $^a$Department of Physics, Aoyama Gakuin University,\\
 Setagaya, Tokyo 157-8572, Japan

\par\vspace{2mm}\par
 $^b$Department of Crystalline Materials Science,
 Nagoya University,\\ Nagoya 464-8603, Japan

\par\vspace{2mm}\par
 $^c$Center for Integrated Research in Science and Engineering,\\
 Nagoya University,
 Nagoya 464-8603, Japan

\end{center}

\par\vspace{4mm}\par
\noindent
Low temperature resistivity of
CMR manganites is investigated.
At the ground state,
 conduction electrons are perfectly spin polarized, which is
called {\em half-metallic}.
From one-magnon scattering processes, it is discussed that the
resistivity of a half metal as a function of temperature
scales as $\rho(T) - \rho(0) \propto T^3$.
We take   (Nd$_{0.8}$Tb$_{0.2}$)$_{0.6}$Sr$_{0.4}$MnO$_3$
as an example to compare theory and experiments.
The result is in a good agreement.


\section{Introduction.}

One of the novel characters of colossal magnetoresistance 
(CMR) manganites is its half-metallic
ground state \cite{Furukawa99}. 
Due to the strong Hund's coupling between localized $t_{2\rm g}$ spins
and itinerant $e_{\rm g}$ electrons, conduction electrons are
perfectly spin polarized at the ferromagnetic ground state \cite{Zener51}.
Half metallic natures of manganites are important to understand
various properties of CMR manganites at low temperatures. 
Especially, spin valve mechanism of the inter-domain tunneling currents
is considered to be
 the origin of the low field magnetoresistance phenomena
in polycrystals \cite{Hwang96} as well as hetero-junctions \cite{Sun96}.

Low temperature resistivity behavior is 
another feature  of half metals.
It has been first pointed out by Kubo and Ohata \cite{Kubo72}
that the perfect spin polarization of conduction electrons
make a qualitative change in the scattering processes of 
charge carriers by magnon-electron interactions.
Half metals belong to a different universality class from
conventional itinerant weak ferromagnets.

Let us discuss the detail of the magnon-electron
scattering process through the temperature dependence of
the resistivity in half metals.
In conventional itinerant weak ferromegnets,
one magnon scattering is the dominant process at low temperatures
to give $T^2$ resistivity \cite{Mannari59}.
However, for half metals, such a scattering process is
prohibited at the ground state since the process involves
low energy propagators of spin-flipped quasi-particles
for the intermediate states.
Based on a rigid band electronic structures of half metals,
Kubo and Ohata  discussed that the most dominant scattering
process is the two-magnon scattering, which gives $T^{4.5}$.

At finite temperatures, however, it is necessary to take into account
the effect of spin fluctuations which breaks down the perfect
 spin polarization. In the absence of spin gaps, magnetization 
deviates from its saturation values as $\delta M \propto T^{3/2}$
in three dimensions. In the strong Hund's coupling limit, 
spin polarization of the conduction electrons are proportional to
the total spin polarization. Namely, at finite temperatures,
half-metallic structure of conduction electrons breaks down.
As a consequence, the rigid band approaches should not be justified.
Taking into account the non rigid band behavior due to
spin fluctuations, one of the author (N.F.) derived \cite{Furukawa99x}
that the most dominant contribution for the low temperature
resistivity is from an unconventional one-magnon scattering processes.
In this case, the  resistivity is proportional to the 
product of following two quantities;
(a) magnon population $\delta M$, and (b) 
density of states of the minority spin quasiparticles
which should also scale as $\delta M$.
As a consequence, we obtain
\begin{equation}
 \rho(T) - \rho(0) \propto (\delta M)^2\propto T^3.
\end{equation}

Experimentally,
it has been reported that the low temperature resistivity
for La$_{1-x}$Sr$_{x}$MnO$_3$ \cite{Furukawa99x} 
as well as Sm$_{0.6}$Sr$_{0.4}$MnO$_3$ scales as $T^3$,
which is another evidence for the half-metallic behaviors of
CMR manganites.
In this paper, we investigate the temperature dependence of the
resistivity in CMR manganite
(Nd,Tb)$_{0.6}$Sr$_{0.4}$MnO$_3$ in detail
as another candidate to investigate the half-metallic nature 
at low temperatures.
From its resistivity as a function of temperature,
we discuss the electronic structures as well as
scattering processes 
in the low temperature region.

\section{Low temperature resistivity in manganites.}

Here we show the low temperature resistivity
of a CMR manganite in the ferromagnetic metal region.
Single crystal samples of (Nd$_{0.8}$Tb$_{0.2}$)$_{0.6}$Sr$_{0.4}$MnO$_3$
are used for the resistivity measurement.
Details of the sample preparation 
have been reported in ref.~\cite{Akimoto00x}.
Curie temperature is estimated as $T_{\rm c} \simeq 200{\rm K}$,
which is substantially decreased from the value for 
La$_{0.6}$Sr$_{0.4}$MnO$_3$, $T_{\rm c} \simeq 400{\rm K}$.
This means that this compound is in the narrow bandwidth region.

\begin{figure}
\epsfxsize=10cm
\hfil\epsfbox{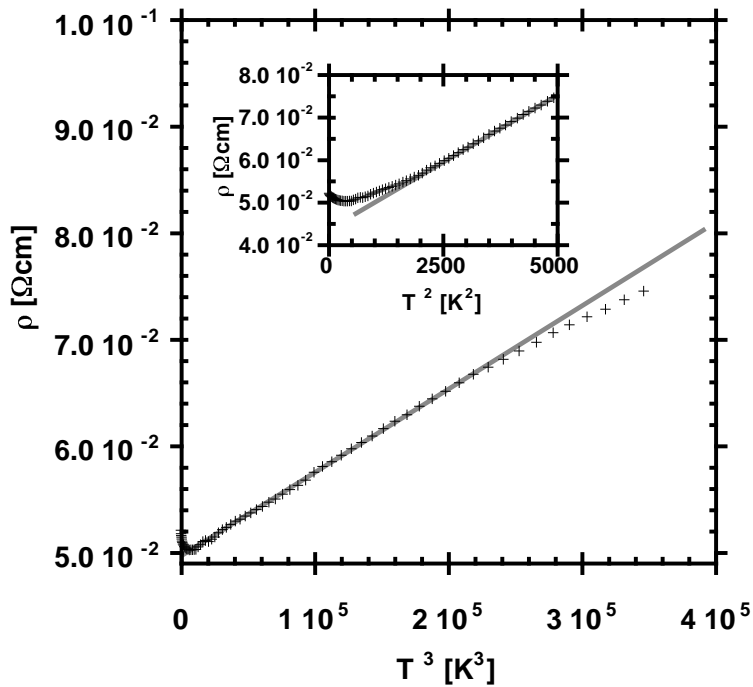}\hfil

\caption{ $T^3$ plot for the low temperature resistivity $\rho(T)$ of
(Nd$_{0.8}$Tb$_{0.2}$)$_{0.6}$Sr$_{0.4}$MnO$_3$. Inset: $T^2$ plot.}
\end{figure}

In Fig.~1 we show the resistivity as a function of
temperature. The data can be fitted in the form
 $\rho(T) - \mbox{const.} \propto T^3$ at the low temperature region.
A crossover  in temperature dependence is observed
at $T^* \simeq 50{\rm K}$, above which the resistivity scales as
 $\rho(T) - \mbox{const.} \propto T^2$ at $T>T^*$.
At the lowest temperature limit we see an upturn of the resistivity
which presumably indicates the localization of carriers,
as also seen in La$_{1-x}$Sr$_{x}$MnO$_3$ near the metal-insulator
transition \cite{Okuda98}. Together with the fact that 
the value of the residual resistivity is large, 
the system seems to be  substantially influenced by the randomness
due to ternary mixture of A-site ions.
Nevertheless, we see $T^3$ dependence of resistivity
which indicates the robustness of the one-magnon
scattering processes in half metals.

\section{Discussion}

Thus (Nd$_{0.8}$Tb$_{0.2}$)$_{0.6}$Sr$_{0.4}$MnO$_3$ show a common
feature in low temperature resistivity
as La$_{1-x}$Sr$_{x}$MnO$_3$ and 
 Sm$_{0.6}$Sr$_{0.4}$MnO$_3$.
In CMR manganites,  phase controls have been performed through the
chemical pressure effects in A-site substitutions. For the above example,
$T_{\rm c}$ varies from $T_{\rm c} \sim 350{\rm K}$ 
((La,Sr)MnO$_3$) to $T_{\rm c} \sim 120{\rm K}$ ((Sm,Sr)MnO$_3$).
This indicates that, once the ferromagnetic metal phases are
stabilized, they share a common feature 
based on the half-metallic structure of conduction electrons.
We also 
note that similar $T^3$ behavior in resistivity is reported for
 CrO$_2$ films \cite{Li99}, which is another candidate for a half metal.
This work was supported by the
Grant-In-Aid for Scientific Research from the
Ministry of Education, Science, Sports and Culture.

\end{document}